\documentclass[11pt,a4paper]{article}

\usepackage[utf8]{inputenc}
\usepackage[T1]{fontenc}
\usepackage{lmodern}
\usepackage{amsmath}
\usepackage{amssymb}
\usepackage{amsthm}
\usepackage{booktabs}
\usepackage{longtable}
\usepackage{array}
\usepackage{calc}
\usepackage{etoolbox}
\usepackage{graphicx}
\usepackage{float}
\usepackage{fvextra}
\usepackage{microtype}
\usepackage[numbers,sort&compress]{natbib}
\usepackage[margin=1in]{geometry}
\usepackage[colorlinks=true,linkcolor=blue,citecolor=blue,urlcolor=blue]{hyperref}
\providecommand{\tightlist}{\setlength{\itemsep}{0pt}\setlength{\parskip}{0pt}}

\makeatletter
\def\maxwidth{\ifdim\Gin@nat@width>\linewidth\linewidth\else\Gin@nat@width\fi}
\def\maxheight{\ifdim\Gin@nat@height>\textheight\textheight\else\Gin@nat@height\fi}
\makeatother
\setkeys{Gin}{width=\maxwidth,height=\maxheight,keepaspectratio}
\DefineVerbatimEnvironment{verbatim}{Verbatim}{breaklines,breakanywhere,fontsize=\small}
\AtBeginEnvironment{longtable}{\small}
\renewcommand{\_}{\textunderscore\allowbreak}
\title{Probabilistic Sensing, Deterministic Authority: Admitting Model-Produced Observations into Sufficiency-Checked Governance Contracts}
\author{Gaston Besanson\thanks{Universidad Torcuato Di Tella}}
\date{}

\begin{document}
\maketitle

\begin{abstract}
When a field that an authority contract needs exists only in unstructured evidence, a model can sense it. We admit the model's output only as an observation record with a score. An admission policy, with thresholds fitted on a held-out split at a declared false-positive ceiling, maps each score to true, false or unknown. Unknown denies. A deterministic, sufficiency-checked contract decides. The probability that sensing changes the verdict is bounded by the sum, over the contract's sensed fields, of the admitted-wrong and unknown rates. This is an instantiation of union-bound reasoning, indexed by the contract. Minimising the estimated bound is a valid cost model for choosing among sufficient contracts. In a registered study on two constructed domains with two sensor families (36,000 model calls), no cell refuted the bound. Deny-to-allow changes from sensing appeared for the first time in this programme: 13 of 21,000 test verdicts, all from 3 contradictory records; each flip in a cell with a registered bound lay under it. Sensing-aware selection picked the lower-exposure contract in 4 of 4 registered tests. Both sensors' scores were informative but not calibrated. Correctness is relative to the declared loss model, candidate representation and reachable states; all domains are constructed.
\end{abstract}

\hypertarget{introduction}{%
\section{Introduction}\label{introduction}}

Papers 1 to 5 of this series govern agent actions with facts read from systems of record. Paper 3 gates the quality of the evidence an agent acts on \citep{besanson2026sarcdq}. Paper 4 composes several pre-action controls on one action \citep{besanson2026onegate}. Paper 5 compiles, from a declared loss model and a declared reachable-state model, the observation contracts that are sufficient to reproduce a verdict, and selects among them by cardinality or declared cost \citep{besanson2026compiling}. Every field those contracts read is recorded: it comes from a system that stores it, with metadata a data-quality check can test. That leaves a gap. Some fields an authority contract needs are written down only in unstructured evidence: a change ticket that says an approval was given, a contract that says where data are stored.

\textbf{What changes.} This paper adds a third kind of input. A model may read the document and \emph{sense} the field. The sensor writes an observation record with a score per candidate value. An admission policy, a pair of thresholds per value fitted on held-out data, maps the score to true, false or unknown. The contract then decides exactly as before.

\textbf{What does not change.} The gate code, the compiler and the invariants of the earlier papers stay as they are. A sensed record is a new record type in the governed buffer; it is never written back to a source system. The invariant this paper adds is one sentence. A model's output enters the gate as an observation with an admission policy. It never carries authority.

\textbf{Contributions.}

\begin{itemize}
\tightlist
\item
  \emph{Architecture (claim C1).} A sensed field enters the gate only as a stamped observation record with provenance, score and admission decision. The model never emits a verdict (Sections 2 and 3).
\item
  \emph{Bound and selection (claims C2 and C3).} Proposition S1 bounds the probability that sensing changes the verdict by the sum, over the contract's sensed fields, of the rates of wrong admission and of unknown. Proposition S2 makes the estimated bound a valid cost model for paper 5's selection, for nested sensed-field sets (Section 4).
\item
  \emph{Registered evidence (claims C4, C5 and C7).} On two constructed domains, with two sensor families and three noise levels, no registered cell refuted the bound; the first deny-to-allow flips of the programme appeared, rarely, and under the registered bound wherever one exists; and sensing-aware selection picked the contract with lower measured exposure in every registered test (Section 7).
\item
  \emph{Calibration (claim C6).} Neither sensor family's scores were calibrated, although both were informative. Thresholds must therefore be fitted, not assumed (Section 7).
\end{itemize}

\textbf{Reader's map.} Section 2 sets out the objects and Figure 1 the pipeline. Section 3 states the invariants. Section 4 gives the propositions and their proof status. Section 5 describes admission and estimation. Section 6 summarises the registered design and Section 7 the results. Sections 8 to 11 cover related work, claims, limitations and reproducibility. Appendix A carries the proofs and Appendix B the verbatim registered materials.

\hypertarget{setting}{%
\section{Setting}\label{setting}}

We inherit paper 5's notation. \(A\) is the set of candidate attributes, \(R\) the set of reachable tuples, and \(g: R \to \{\mathit{allow}, \mathit{deny}\}\) the declared verdict. A contract \(C \subseteq A\) is \emph{sufficient} if \(g\) is determined by \(C\) on \(R\): two reachable tuples that agree on \(C\) get the same verdict.

\textbf{New objects.} The contract is partitioned as \(C = F_r \cup F_s\). Recorded fields \(F_r\) are read from systems of record. Sensed fields \(F_s\) are produced by sensors. A sensor \(S_i\) for field \(i\) maps an evidence document \(d\) to a score \(s \in [0,1]\) for each candidate value of the field, one value against the rest. An admission policy \(A_i\) maps each value's score to true, false or unknown with two thresholds, \(\tau^{\mathrm{true}}\) and \(\tau^{\mathrm{false}}\), fitted at a false-positive target \(\varepsilon\). The field's admitted value is \(v\) if exactly one value maps to true; otherwise the field is unknown, and unknown denies.

\textbf{The sensed record.} Each admission is written as a record with the fields \texttt{request\_id}, \texttt{resource\_id}, \texttt{field}, \texttt{value}, \texttt{score}, \texttt{admission}, \texttt{admission\_policy\_version}, the sensor's identifier and version, the SHA-256 of the source document and \texttt{sensed\_at}. It carries metadata, so the data-quality predicates of the earlier papers apply to it: freshness of sensing, the sensor version pin, the admission policy version pin, and lineage to the document, the request and the resource. A record whose stamp does not match the request being decided is rejected and the field is unknown.

\textbf{Evaluation from observations.} Given an observed value for every field of \(C\) (recorded values as recorded, sensed values as admitted, derived values computed as in Section 3), the verdict is \(g(t')\) for any \(t' \in R\) that agrees with the observations on \(C\). Sufficiency makes this well defined. If a sensed field is unknown, or no reachable tuple agrees, the verdict is deny.

\textbf{Exposure.} For a request with true tuple \(t\), exposure is the event that the verdict from recorded and sensed values differs from the verdict from recorded and true values. Throughout, recorded fields are taken at their true values, and every derivation is truth-preserving: a true assertion derives the true value of the property it determines. The reference verdict is therefore \(g(t)\). A change from deny to allow is \emph{unsafe}; a change from allow to deny is \emph{fail-closed}. For sensed field \(i\) and a distribution over requests, \(e_i\) is the probability that the field is admitted with a wrong value and \(u_i\) the probability that it is unknown.

\begin{figure}
\centering
\includegraphics{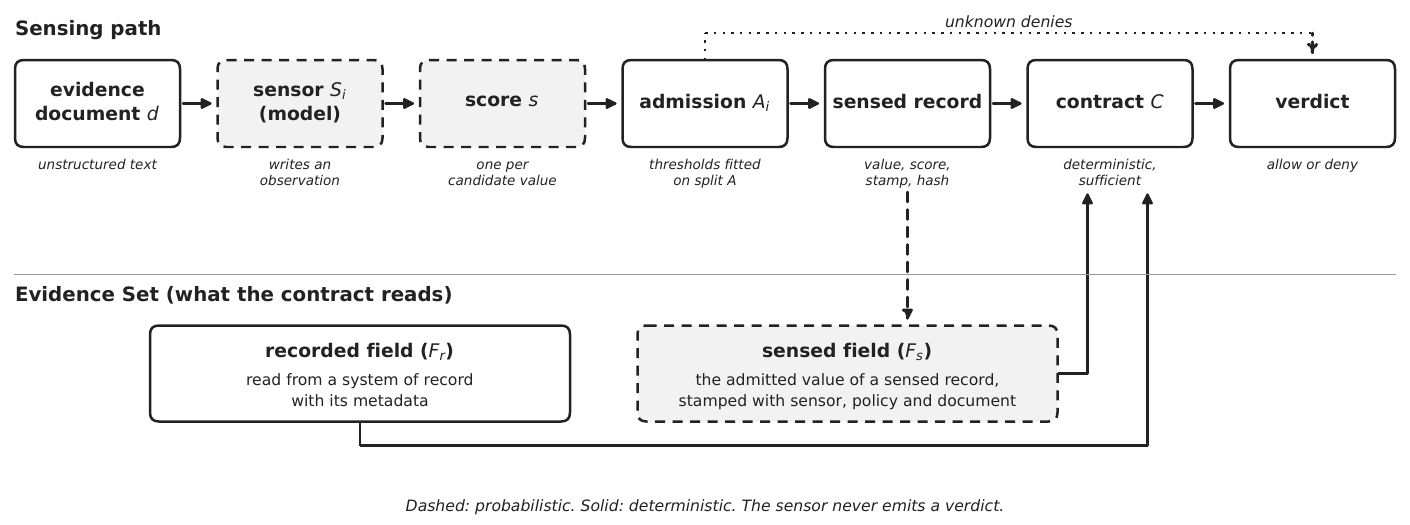}
\caption{The sensing pipeline. The top row is the sensing path from document to verdict; dashed boxes are probabilistic and solid boxes deterministic. The bottom row is the Evidence Set line: the contract reads a recorded field and a sensed field side by side, and the sensed field reaches it only as the admitted value of a stamped record.}
\end{figure}

\hypertarget{invariants}{%
\section{Invariants}\label{invariants}}

\textbf{I1, carried from paper 3.} Validity is never elicited from a model. The sensor writes an observation; the deterministic contract computes the verdict. The admission policy is a threshold with abstention per field, never a weighted combination across fields.

This paper makes one exception to the earlier practice of reading every field from a record, and states it precisely. A sensed field's value may come from a model, but only as an admitted \emph{assertion}: what the document states. Nothing else about the field comes from the model.

\textbf{I2.} A sensor is asked only what the document asserts: whether an approval is recorded or not recorded, the date the document gives, where it says the data are stored, and which branch or environment it names. Validity, scope and expiry are never sensed. They are computed deterministically from recorded fields plus the admitted assertion. On the CH-C1 domain, for example, the sensed field is \texttt{approval\_assertion}, and \texttt{approval\_token} is derived from it: absent if the assertion is \texttt{not\_recorded}; valid if the asserted date lies within 30 days before the recorded request date; expired otherwise.

I2 answers the obvious objection to letting a model near an authority gate. The model is not asked whether an approval is valid, current or sufficient. It is asked what a document says, which is a reading task with a checkable label. The judgement that turns the reading into a permission stays in code that can be inspected and tested.

\hypertarget{propositions}{%
\section{Propositions}\label{propositions}}

\textbf{Proposition S1 (union bound).} For a contract with sensed-field set \(F_s\), \[P(\text{verdict change}) \le \sum_{i \in F_s} (e_i + u_i), \qquad P(\text{deny} \to \text{allow}) \le \sum_{i \in F_s} e_i .\] \emph{Sketch.} If every sensed field is admitted with its true value, the observations agree with the true tuple on \(C\), so sufficiency gives the true verdict. A verdict change therefore requires at least one sensed field that is wrong or unknown. A change to allow also requires that no field is unknown, because unknown denies, so it requires at least one wrongly admitted field. Both inequalities follow from the union bound. The argument is pointwise: the event inclusion holds at every pair of a true tuple and an observation vector, and both sides are linear in the joint distribution, so the inequalities hold for every distribution, including any dependence between fields. S1 is an instantiation of union-bound reasoning over program events \citep{barthe2016program}, indexed by the contract's sensed fields. It is not a new inequality.

\textbf{Proposition S2 (monotonicity for nested sets).} For two sufficient contracts whose sensed-field sets are nested, \(F_s' \subseteq F_s\), read by the same sensors on the same documents through the same admission policy and the same derivation, the S1 bound of \(F_s'\) does not exceed that of \(F_s\). The selection objective is the \emph{estimated} bound \(\sum_{i \in F_s} (\hat e_i + \hat u_i)\) of Section 5. It fits paper 5's observation-cost interface, so no new solver is needed. The interface charges each contract property the estimated terms of the sensed field it is derived from (\texttt{approval\_assertion} for \texttt{approval\_token}), plus a floor kept below half of every gap between distinct estimated bounds, so the floor breaks only exact ties. No claim is made for sets that are not nested. The S2 checker records why: with the per-field rates below, a rule that prefers fewer sensed fields picks the one-field set, while the registered objective picks the two-field set, whose bound is a hundred times lower.

\begin{longtable}[]{@{}
  >{\raggedright\arraybackslash}p{(\columnwidth - 8\tabcolsep) * \real{0.1892}}
  >{\raggedright\arraybackslash}p{(\columnwidth - 8\tabcolsep) * \real{0.1351}}
  >{\raggedright\arraybackslash}p{(\columnwidth - 8\tabcolsep) * \real{0.2432}}
  >{\raggedright\arraybackslash}p{(\columnwidth - 8\tabcolsep) * \real{0.1351}}
  >{\raggedright\arraybackslash}p{(\columnwidth - 8\tabcolsep) * \real{0.2973}}@{}}
\toprule\noalign{}
\begin{minipage}[b]{\linewidth}\raggedright
sensed-field set
\end{minipage} & \begin{minipage}[b]{\linewidth}\raggedright
fields
\end{minipage} & \begin{minipage}[b]{\linewidth}\raggedright
\(e\) per field
\end{minipage} & \begin{minipage}[b]{\linewidth}\raggedright
S1 bound
\end{minipage} & \begin{minipage}[b]{\linewidth}\raggedright
picked by
\end{minipage} \\
\midrule\noalign{}
\endhead
\bottomrule\noalign{}
\endlastfoot
one field & a & 0.2, 0.0 & 0.2 & fewest sensed fields \\
two fields & b, c & 0.001, 0.0; 0.001, 0.0 & 0.002 & registered objective \\
\end{longtable}

The rate columns list \(e\) and then \(u\) for each field, as the checker records them.

\textbf{Proposition S3 (fail-closed).} Call a sensing configuration \emph{deny-ward} if, for every reachable tuple \(t\) with \(g(t) = \mathit{deny}\) and every observation vector that agrees with \(t\) on \(F_r\) and whose sensed fields are each either true or admitted with any wrong value, jointly, the contract evaluates to deny. If unknown routes to deny and the configuration is deny-ward, then deny-to-allow changes have probability zero. The condition is global over \(R\) and over joint errors, not field by field. Section 7 evaluates it exhaustively for every registered configuration; where it fails, S3 does not apply, and zero observed deny-to-allow changes are an empirical result only.

\textbf{N6 (negative witnesses).} Two constructions mark the edges of S1 and S3. In witness (a), two binary fields with disjoint wrong admissions attain the S1 deny-to-allow bound exactly. In witness (b), an AND contract with true values (false, false) meets a sensor that admits both fields wrongly together with probability \(q\) and never one alone. Every single-field error is harmless, so a field-by-field direction test calls the configuration deny-ward, yet the deny-to-allow probability is \(q\). The first registration of this paper used such a directional bound; witness (b) defeats it. The external review of the novelty fence raised the point, and the re-registration from \texttt{prereg-p6-v1} to \texttt{prereg-p6-v1.1} replaced the directional term with S1 and made S3's condition global.

Each statement carries the proof-status tag recorded in \texttt{proof\_status.json}. A tag of machine-checked means a checker holds on exactly the statement's scope; checked-scope-only means a checker holds on a narrower scope; pending-human-review means no checker covers it.

\begin{longtable}[]{@{}
  >{\raggedright\arraybackslash}p{(\columnwidth - 6\tabcolsep) * \real{0.3191}}
  >{\raggedright\arraybackslash}p{(\columnwidth - 6\tabcolsep) * \real{0.2766}}
  >{\raggedright\arraybackslash}p{(\columnwidth - 6\tabcolsep) * \real{0.1277}}
  >{\raggedright\arraybackslash}p{(\columnwidth - 6\tabcolsep) * \real{0.2766}}@{}}
\toprule\noalign{}
\begin{minipage}[b]{\linewidth}\raggedright
statement
\end{minipage} & \begin{minipage}[b]{\linewidth}\raggedright
scope
\end{minipage} & \begin{minipage}[b]{\linewidth}\raggedright
checker
\end{minipage} & \begin{minipage}[b]{\linewidth}\raggedright
tag
\end{minipage} \\
\midrule\noalign{}
\endhead
\bottomrule\noalign{}
\endlastfoot
S1 on the finite model families & finite-models:B12,B3,T2 & \texttt{s1\_check} & machine-checked \\
S1, general & general & \texttt{s1\_check} & checked-scope-only \\
S2 on the finite model families, nested sets & finite-models:B12,B3,T2;nested-sets-only & \texttt{s2\_check} & machine-checked \\
S2, general & general & \texttt{s2\_check} & checked-scope-only \\
S3 on the finite model families & finite-models:B12,B3,T2 & \texttt{s3\_check} & machine-checked \\
S3, general & general & \texttt{s3\_check} & checked-scope-only \\
N6, both witnesses & witness & \texttt{n6\_witness} & machine-checked \\
written proof of S1 (Appendix A) & general & none & pending-human-review \\
written proof of S2 (Appendix A) & general & none & pending-human-review \\
written proof of S3 (Appendix A) & general & none & pending-human-review \\
\end{longtable}

\hypertarget{admission-policy-and-estimation}{%
\section{Admission policy and estimation}\label{admission-policy-and-estimation}}

Each experiment's validation split is divided once, before any model call, into split A and split B of 150 items each (seed 20261107). The two halves have different jobs.

\textbf{Split A fits the thresholds against the assertion label.} For each sensor, field and candidate value \(v\), \(\tau_v^{\mathrm{true}}\) is the lowest candidate threshold at which at most \(\varepsilon\) of split-A records whose assertion label for \(v\) is zero score at or above it; \(\tau_v^{\mathrm{false}}\) is the highest at which at most \(\varepsilon\) of records whose label is one score at or below it. The assertion label says what the document states, so the thresholds measure reading fidelity. The ceiling \(\varepsilon\) is 1\%, with 5\% as an ablation. The ceiling is a fitting target on split A, not a property of the test split.

\textbf{Split B estimates the error rates against the truth label.} With the thresholds frozen, \(\hat e_i\) and \(\hat u_i\) are one-sided 95\% Clopper-Pearson upper bounds \citep{clopper1934use} on the split-B rates of wrong admission and of unknown, per sensor, field and noise level. The truth label is the item's true value. A faithful reading of a contradictory document can therefore count as a wrong admission. The registered bounds per cell are \(B^{+} = \sum_{i \in F_s} \hat e_i\) for deny to allow and \(B = \sum_{i \in F_s} (\hat e_i + \hat u_i)\) for any change. Thresholds and estimates were written to a committed file before any test-split call.

\textbf{The floor.} A Clopper-Pearson upper bound is never zero. With no error among 150 split-B records, the bound for one field is 0.0198. With two sensed fields, a sensor that made no split-B error still gets \(B^{+}\) = 0.0395 and \(B\) = 0.0791.

\textbf{From S1 to the estimated bounds.} S1 is valid, not tight. The estimated bounds \(B^{+}\) and \(B\) replace S1's population rates with split-B upper limits, and that step needs two assumptions. First, \emph{transfer}: the sensors' error and unknown rates on the test documents are no higher than on split B, although test documents are drawn from a disjoint phrase bank. Second, \emph{simultaneity}: each limit is a marginal 95\% limit, so a sum of them carries no joint 95\% level, and neither does a family of cells. Section 7 reports, beside each registered bound, a Bonferroni-adjusted simultaneous version, in which each of the \(m\) summed limits is taken at level \(0.05/m\).

\hypertarget{experimental-design}{%
\section{Experimental design}\label{experimental-design}}

Everything below was registered before any experiment code existed. The first registration is \texttt{prereg/p6-v1.md} (tag \texttt{prereg-p6-v1}, commit \texttt{ddcddf4}). After an external review of the novelty fence, the author re-registered as \texttt{prereg/p6-v1.1.md} (tag \texttt{prereg-p6-v1.1}, commit \texttt{2fac5a4}). The review is committed unedited under \texttt{review-secondary/} and was not applied to any file.

\textbf{Sensors.} Two sensor families read the same documents. Jev (System One, model \texttt{jev-latest}, returned as \texttt{jev-1.13.0}) is a typed probabilistic model from TypeSafe; it receives one question per candidate value. Claude Haiku 4.5 (\texttt{claude-haiku-4-5-20251001}) is a generative language model from Anthropic; it receives one prompt per field and returns a JSON object of probabilities. TypeSafe and Anthropic appear here only as the vendors of the tested configurations. The questions and the prompt are in Appendix B.

\textbf{E0, motivating evidence.} The earlier probes jev-v1, jev-v2 and jev-v3 are quoted from tag \texttt{jev-probes-final} (commit \texttt{e1fc75b}). Nothing is re-run.

\textbf{E1, the bound on CH-C1.} Paper 5's 35-property domain, with contract K, the first minimum-cardinality contract of paper 5's CH-C1 checker. A registered rule picks the two document-backed fields of K with the highest declared observation cost: \texttt{approval\_token}, sensed as \texttt{approval\_assertion} over \texttt{not\_recorded} and six registered dates, and \texttt{data\_residency\_region}. All other fields are recorded at their true values. The population is 27,000 reachable tuples. From it, 1,000 items were drawn (seed 20261101), with a validation split of 300 (seed 20261102) and a test split of 700.

\textbf{E2, the substitution test on CH-B1.} Paper 5's code and cloud domain, with jev-v2's items and split. CH-B1 has two seven-attribute reducts, \(R_{\mathrm{branch}}\) and \(R_{\mathrm{env}}\). In arm 1, \texttt{branch} is sensed and \texttt{environment} recorded, so \(R_{\mathrm{env}}\) senses only the approval field and \(R_{\mathrm{branch}}\) senses two fields. Arm 2 is the mirror image. In each arm the sensed-field sets are nested, so S2 applies. The registered rule picks the reduct with the lower estimated bound \(B\) on split B at 30\% noise, before any test-split call. Both reducts are evaluated on the same readings, so every test item gives a pair of outcomes.

\textbf{E3 to E5.} E3 re-reads the E1 caches at ceilings of 1\% and 5\%, and with unknown routed to deny or to escalation. E4 runs the two N6 witnesses through the pipeline on 10,000 synthetic items each, with no model. E5 measures calibration against the assertion label.

\textbf{Documents and labels.} Each item gets one synthetic record per noise level, generated from a seeded generator and never edited. The record carries no property value except the sensed statements. Validation and test items draw their statements from disjoint phrase banks. At noise levels 0\%, 10\% and 30\%, each sensed statement is independently perturbed: \emph{missing} removes it, and \emph{contradictory} keeps it and appends a comment asserting a different value. Noise is nested across levels. Each item and field has two labels. The \emph{truth label} is the true value. The \emph{assertion label} marks every value the record states, so a contradictory record asserts two values and a missing one asserts none.

\textbf{Run discipline.} The order was registered: E1 before E2; within each, all validation records, then the arm health check, then thresholds written and committed, then the test split. The arm health rule stops a run before any further call if more than 2 invalid replies occur in the first 20 validation records of a sensor, or if more than 5\% of its validation replies are invalid. A cap of USD 60 covered E1 to E5. Every raw response was appended to a cache before parsing. Every number in the results is a named slot filled from the caches by one command.

\hypertarget{results}{%
\section{Results}\label{results}}

\textbf{Runs.} Both experiments completed: 36,000 calls (12,000 in E1 and 24,000 in E2), 0 invalid replies, USD 8.26 in total. Returned model strings: claude-haiku-4-5-20251001, jev-1.13.0 (E1); claude-haiku-4-5-20251001, jev-1.13.0 (E2).

\hypertarget{e0-the-earlier-probes}{%
\subsection{E0: the earlier probes}\label{e0-the-earlier-probes}}

The earlier probes are quoted from tag \texttt{jev-probes-final}. Across two sensors, three noise levels and 4,200 test verdicts on one sensed field, no sensing error turned deny into allow (0 of 4,200); every change was fail-closed. Calibration was refuted for Jev in jev-v1 (Spiegelhalter \(Z\) = -15.0268, Holm-adjusted \(p\) = 1.47e-50, verdict: refuted). Known issue KI-1 is part of the record. In jev-v2, Haiku wrapped every reply in a code fence and the registered parser scored 3000 of 3000 replies invalid, so that arm denied everything. jev-v3 re-ran the arm under a registered fix and an arm health check, with 0 invalid replies. The arm health rule of this paper comes from that record.

\hypertarget{e1-the-bound-on-ch-c1}{%
\subsection{E1: the bound on CH-C1}\label{e1-the-bound-on-ch-c1}}

Each cell is a sensor at a noise level, on the 700 test items. Rates carry Clopper-Pearson 95\% intervals. \(B^{+}\) and \(B\) are the split-B bounds of Section 5.

\begin{longtable}[]{@{}
  >{\raggedright\arraybackslash}p{(\columnwidth - 16\tabcolsep) * \real{0.0745}}
  >{\raggedright\arraybackslash}p{(\columnwidth - 16\tabcolsep) * \real{0.0638}}
  >{\raggedright\arraybackslash}p{(\columnwidth - 16\tabcolsep) * \real{0.0851}}
  >{\raggedright\arraybackslash}p{(\columnwidth - 16\tabcolsep) * \real{0.1702}}
  >{\raggedright\arraybackslash}p{(\columnwidth - 16\tabcolsep) * \real{0.0851}}
  >{\raggedright\arraybackslash}p{(\columnwidth - 16\tabcolsep) * \real{0.1277}}
  >{\raggedright\arraybackslash}p{(\columnwidth - 16\tabcolsep) * \real{0.1702}}
  >{\raggedright\arraybackslash}p{(\columnwidth - 16\tabcolsep) * \real{0.1277}}
  >{\raggedright\arraybackslash}p{(\columnwidth - 16\tabcolsep) * \real{0.0957}}@{}}
\toprule\noalign{}
\begin{minipage}[b]{\linewidth}\raggedright
sensor
\end{minipage} & \begin{minipage}[b]{\linewidth}\raggedright
noise
\end{minipage} & \begin{minipage}[b]{\linewidth}\raggedright
unsafe (\(k/n\))
\end{minipage} & \begin{minipage}[b]{\linewidth}\raggedright
unsafe 95\% CI
\end{minipage} & \begin{minipage}[b]{\linewidth}\raggedright
\(B^{+}\)
\end{minipage} & \begin{minipage}[b]{\linewidth}\raggedright
change rate (\(k/n\))
\end{minipage} & \begin{minipage}[b]{\linewidth}\raggedright
change 95\% CI
\end{minipage} & \begin{minipage}[b]{\linewidth}\raggedright
\(B\)
\end{minipage} & \begin{minipage}[b]{\linewidth}\raggedright
\end{minipage} \\
\midrule\noalign{}
\endhead
\bottomrule\noalign{}
\endlastfoot
Jev & 0\% & 0/700 & {[}0.0000, 0.0053{]} & 0.0395 & 0.0000 (0/700) & {[}0.0000, 0.0053{]} & 0.0791 & \\
Jev & 10\% & 0/700 & {[}0.0000, 0.0053{]} & 0.0611 & 0.0429 (30/700) & {[}0.0291, 0.0606{]} & 0.3045 & \\
Jev & 30\% & 1/700 & {[}0.0000, 0.0079{]} & 0.1056 & 0.1071 (75/700) & {[}0.0852, 0.1324{]} & 0.7837 & \\
Haiku & 0\% & 0/700 & {[}0.0000, 0.0053{]} & 0.0395 & 0.0643 (45/700) & {[}0.0473, 0.0851{]} & 0.4268 & \\
Haiku & 10\% & 0/700 & {[}0.0000, 0.0053{]} & 0.0798 & 0.0943 (66/700) & {[}0.0737, 0.1184{]} & 0.6324 & \\
Haiku & 30\% & 0/700 & {[}0.0000, 0.0053{]} & 0.0707 & 0.1471 (103/700) & {[}0.1217, 0.1756{]} & 1.0761 vacuous & \\
\end{longtable}

A bound greater than one is marked vacuous: it holds for any sensor. Vacuous bounds in E1: \(B\) for Haiku at 30\% noise. The floor of Section 5 applies to every cell: with no split-B error among 150 records per field, the bound is 0.0198 for \texttt{approval\_assertion} and 0.0198 for \texttt{data\_residency\_region}, so \(B^{+}\) cannot fall below 0.0395 and \(B\) below 0.0791.

No cell's unsafe rate exceeded its \(B^{+}\) (H1, Section 7.8). The comparison rests on the two assumptions of Section 5. The registered bounds and their simultaneous versions are:

\begin{longtable}[]{@{}
  >{\raggedright\arraybackslash}p{(\columnwidth - 10\tabcolsep) * \real{0.1111}}
  >{\raggedright\arraybackslash}p{(\columnwidth - 10\tabcolsep) * \real{0.0972}}
  >{\raggedright\arraybackslash}p{(\columnwidth - 10\tabcolsep) * \real{0.1250}}
  >{\raggedright\arraybackslash}p{(\columnwidth - 10\tabcolsep) * \real{0.2778}}
  >{\raggedright\arraybackslash}p{(\columnwidth - 10\tabcolsep) * \real{0.1111}}
  >{\raggedright\arraybackslash}p{(\columnwidth - 10\tabcolsep) * \real{0.2778}}@{}}
\toprule\noalign{}
\begin{minipage}[b]{\linewidth}\raggedright
sensor
\end{minipage} & \begin{minipage}[b]{\linewidth}\raggedright
noise
\end{minipage} & \begin{minipage}[b]{\linewidth}\raggedright
\(B^{+}\)
\end{minipage} & \begin{minipage}[b]{\linewidth}\raggedright
\(B^{+}\) simultaneous
\end{minipage} & \begin{minipage}[b]{\linewidth}\raggedright
\(B\)
\end{minipage} & \begin{minipage}[b]{\linewidth}\raggedright
\(B\) simultaneous
\end{minipage} \\
\midrule\noalign{}
\endhead
\bottomrule\noalign{}
\endlastfoot
Jev & 0\% & 0.0395 & 0.0486 & 0.0791 & 0.1152 \\
Jev & 10\% & 0.0611 & 0.0716 & 0.3045 & 0.3596 \\
Jev & 30\% & 0.1056 & 0.1181 & 0.7837 & 0.8550 \\
Haiku & 0\% & 0.0395 & 0.0486 & 0.4268 & 0.4778 \\
Haiku & 10\% & 0.0798 & 0.0912 & 0.6324 & 0.6981 \\
Haiku & 30\% & 0.0707 & 0.0816 & 1.0761 & 1.1456 vacuous \\
\end{longtable}

The simultaneous versions are descriptive; the registered test uses the marginal ones. Every \(B^{+}\) is at least 0.0395, while the largest unsafe rate observed in any cell is 0.0014. Where verdicts changed at all, \(B\) was 6.6 to 7.3 times the observed change rate.

\textbf{The S3 condition.} S3 needs a global property: every reachable deny tuple stays deny under every joint wrong reading of the sensed fields. Zero observed flips cannot establish it, so it is evaluated directly, exhaustively over the pinned reachable sets (\texttt{experiments/denyward\_p6.py}). The condition holds in 0 of the 5 registered configurations.

\begin{longtable}[]{@{}
  >{\raggedright\arraybackslash}p{(\columnwidth - 10\tabcolsep) * \real{0.0870}}
  >{\raggedright\arraybackslash}p{(\columnwidth - 10\tabcolsep) * \real{0.1304}}
  >{\raggedright\arraybackslash}p{(\columnwidth - 10\tabcolsep) * \real{0.0783}}
  >{\raggedright\arraybackslash}p{(\columnwidth - 10\tabcolsep) * \real{0.0957}}
  >{\raggedright\arraybackslash}p{(\columnwidth - 10\tabcolsep) * \real{0.2000}}
  >{\raggedright\arraybackslash}p{(\columnwidth - 10\tabcolsep) * \real{0.4087}}@{}}
\toprule\noalign{}
\begin{minipage}[b]{\linewidth}\raggedright
contract
\end{minipage} & \begin{minipage}[b]{\linewidth}\raggedright
sensed fields
\end{minipage} & \begin{minipage}[b]{\linewidth}\raggedright
used in
\end{minipage} & \begin{minipage}[b]{\linewidth}\raggedright
deny-ward
\end{minipage} & \begin{minipage}[b]{\linewidth}\raggedright
reachable deny tuples
\end{minipage} & \begin{minipage}[b]{\linewidth}\raggedright
deny tuples some joint reading turns to allow
\end{minipage} \\
\midrule\noalign{}
\endhead
\bottomrule\noalign{}
\endlastfoot
K & approval, residency & E1 & no & 20040 & 4029 \\
\(R_{\mathrm{branch}}\) & approval & E2 arm 2 & no & 8568 & 216 \\
\(R_{\mathrm{branch}}\) & approval, branch & E2 arm 1 & no & 8568 & 216 \\
\(R_{\mathrm{env}}\) & approval & E2 arm 1 & no & 8568 & 216 \\
\(R_{\mathrm{env}}\) & approval, environment & E2 arm 2 & no & 8568 & 216 \\
\end{longtable}

S3 therefore does not apply to any registered configuration. Every cell without a deny-to-allow change is an empirical observation, not an instance of S3.

Per field, on the test split, the rates of wrong admission and of unknown were:

\begin{longtable}[]{@{}
  >{\raggedright\arraybackslash}p{(\columnwidth - 10\tabcolsep) * \real{0.1351}}
  >{\raggedright\arraybackslash}p{(\columnwidth - 10\tabcolsep) * \real{0.1081}}
  >{\raggedright\arraybackslash}p{(\columnwidth - 10\tabcolsep) * \real{0.1892}}
  >{\raggedright\arraybackslash}p{(\columnwidth - 10\tabcolsep) * \real{0.1892}}
  >{\raggedright\arraybackslash}p{(\columnwidth - 10\tabcolsep) * \real{0.1892}}
  >{\raggedright\arraybackslash}p{(\columnwidth - 10\tabcolsep) * \real{0.1892}}@{}}
\toprule\noalign{}
\begin{minipage}[b]{\linewidth}\raggedright
sensor
\end{minipage} & \begin{minipage}[b]{\linewidth}\raggedright
noise
\end{minipage} & \begin{minipage}[b]{\linewidth}\raggedright
approval: wrong
\end{minipage} & \begin{minipage}[b]{\linewidth}\raggedright
approval: unknown
\end{minipage} & \begin{minipage}[b]{\linewidth}\raggedright
residency: wrong
\end{minipage} & \begin{minipage}[b]{\linewidth}\raggedright
residency: unknown
\end{minipage} \\
\midrule\noalign{}
\endhead
\bottomrule\noalign{}
\endlastfoot
Jev & 0\% & 0.0000 & 0.0000 & 0.0000 & 0.0000 \\
Jev & 10\% & 0.0114 & 0.0671 & 0.0100 & 0.0929 \\
Jev & 30\% & 0.0514 & 0.1714 & 0.0186 & 0.2700 \\
Haiku & 0\% & 0.0000 & 0.2443 & 0.0000 & 0.0000 \\
Haiku & 10\% & 0.0171 & 0.2943 & 0.0029 & 0.1086 \\
Haiku & 30\% & 0.0400 & 0.4014 & 0.0100 & 0.3071 \\
\end{longtable}

Haiku's unknown rate on the approval field is high even without noise. The thresholds explain it. On split A, no candidate threshold for Haiku's \texttt{not\_recorded} value met the 1\% ceiling in either direction: its fitted \(\tau^{\mathrm{true}}\) is ``never true'' and its \(\tau^{\mathrm{false}}\) is ``never false'' (Appendix B). That value can never be admitted, so Haiku cannot read a record that asserts no approval correctly: the field is unknown, or wrong if a date value is admitted instead. At 0\% noise its wrong rate on this field was 0.0000, so the admission policy turned a poorly separated score into abstention, and abstention denies.

\hypertarget{the-first-flips}{%
\subsection{The first flips}\label{the-first-flips}}

This programme had recorded no deny-to-allow change from sensing before this paper. E1 and E2 together recorded 13 of 21,000 test verdicts: 1 of 4,200 in E1 (Jev at 30\% noise) and 12 of 16,800 in E2. No cell had more than 1 flip in 700. Registered bounds exist for every E1 cell and for E2 at 30\% noise only, where the selection bound was estimated. The 9 flips in those cells each lay under their registered bound. The other 4 flips are in E2 at 10\% noise, where no bound was registered; Section 7.4 compares them with post hoc bounds.

The count overstates the number of independent events. The item-level record (\texttt{results/p6-flips.json}, written from the caches by \texttt{python\ -m\ experiments.flips\_p6}) traces all 13 flips to 3 records: item 6042 in E1, item 1595 in E2 arm 1 and item 2312 in E2 arm 2. Every flip was on a record whose approval statement had been made contradictory, and in every flip the approval field was the only wrongly admitted field. In E2, both sensors flipped on every such record, and under both reducts in every case, because both reducts read the same approval field. The E1 record shows the mechanism. Its true approval had been recorded but had expired at the request date, so the true verdict was deny. The appended comment said the approval log was empty. Jev admitted \texttt{not\_recorded}, the derived token became absent, and contract K allows that combination.

\emph{Interpretation.} The flips are not misreadings of a clear document. Each record stated two incompatible things about the approval, and the sensors admitted the statement in the appended comment. The truth label counts that as a wrong admission, as registered: exposure is measured against what is true, not against what the document says. The paired flips in E2 are one event per record, seen through two reducts and two sensors, not 12 independent failures. Both reducts read the same approval field, so neither choice of reduct can avoid this failure mode.

\hypertarget{e2-substitution-test-on-ch-b1}{%
\subsection{E2: substitution test on CH-B1}\label{e2-substitution-test-on-ch-b1}}

The picks were fixed on split B at 30\% noise, before any test-split call. In both arms and for both sensors, the rule picked the reduct that senses only the approval field. Paper 5's own choice for CH-B1, its minimum-cost contract under the declared observation costs at the pinned commit, is R\_branch in both arms. Sensing-aware selection therefore changes paper 5's choice in arm 1 and coincides with it in arm 2.

\begin{longtable}[]{@{}
  >{\raggedright\arraybackslash}p{(\columnwidth - 10\tabcolsep) * \real{0.1364}}
  >{\raggedright\arraybackslash}p{(\columnwidth - 10\tabcolsep) * \real{0.0909}}
  >{\raggedright\arraybackslash}p{(\columnwidth - 10\tabcolsep) * \real{0.2500}}
  >{\raggedright\arraybackslash}p{(\columnwidth - 10\tabcolsep) * \real{0.2500}}
  >{\raggedright\arraybackslash}p{(\columnwidth - 10\tabcolsep) * \real{0.1364}}
  >{\raggedright\arraybackslash}p{(\columnwidth - 10\tabcolsep) * \real{0.1364}}@{}}
\toprule\noalign{}
\begin{minipage}[b]{\linewidth}\raggedright
sensor
\end{minipage} & \begin{minipage}[b]{\linewidth}\raggedright
arm
\end{minipage} & \begin{minipage}[b]{\linewidth}\raggedright
estimated bound, \(R_{\mathrm{branch}}\)
\end{minipage} & \begin{minipage}[b]{\linewidth}\raggedright
estimated bound, \(R_{\mathrm{env}}\)
\end{minipage} & \begin{minipage}[b]{\linewidth}\raggedright
picked
\end{minipage} & \begin{minipage}[b]{\linewidth}\raggedright
paper 5's choice
\end{minipage} \\
\midrule\noalign{}
\endhead
\bottomrule\noalign{}
\endlastfoot
Jev & arm 1 & 0.7504 & 0.3171 & R\_env & R\_branch \\
Jev & arm 2 & 0.3170 & 0.7157 & R\_branch & R\_branch \\
Haiku & arm 1 & 0.9182 & 0.4837 & R\_env & R\_branch \\
Haiku & arm 2 & 0.5476 & 0.9884 & R\_branch & R\_branch \\
\end{longtable}

On the test split, the verdict-change and unsafe rates were:

\begin{longtable}[]{@{}
  >{\raggedright\arraybackslash}p{(\columnwidth - 12\tabcolsep) * \real{0.0921}}
  >{\raggedright\arraybackslash}p{(\columnwidth - 12\tabcolsep) * \real{0.0921}}
  >{\raggedright\arraybackslash}p{(\columnwidth - 12\tabcolsep) * \real{0.0789}}
  >{\raggedright\arraybackslash}p{(\columnwidth - 12\tabcolsep) * \real{0.2368}}
  >{\raggedright\arraybackslash}p{(\columnwidth - 12\tabcolsep) * \real{0.2368}}
  >{\raggedright\arraybackslash}p{(\columnwidth - 12\tabcolsep) * \real{0.1316}}
  >{\raggedright\arraybackslash}p{(\columnwidth - 12\tabcolsep) * \real{0.1316}}@{}}
\toprule\noalign{}
\begin{minipage}[b]{\linewidth}\raggedright
sensor
\end{minipage} & \begin{minipage}[b]{\linewidth}\raggedright
arm
\end{minipage} & \begin{minipage}[b]{\linewidth}\raggedright
noise
\end{minipage} & \begin{minipage}[b]{\linewidth}\raggedright
picked: change
\end{minipage} & \begin{minipage}[b]{\linewidth}\raggedright
other: change
\end{minipage} & \begin{minipage}[b]{\linewidth}\raggedright
picked: unsafe
\end{minipage} & \begin{minipage}[b]{\linewidth}\raggedright
other: unsafe
\end{minipage} \\
\midrule\noalign{}
\endhead
\bottomrule\noalign{}
\endlastfoot
Jev & arm 1 & 0\% & 0.0000 (0) & 0.0000 (0) & 0 & 0 \\
Jev & arm 1 & 10\% & 0.0286 (20) & 0.0514 (36) & 0 & 0 \\
Jev & arm 1 & 30\% & 0.0943 (66) & 0.1943 (136) & 1 & 1 \\
Jev & arm 2 & 0\% & 0.0000 (0) & 0.0000 (0) & 0 & 0 \\
Jev & arm 2 & 10\% & 0.0343 (24) & 0.0714 (50) & 1 & 1 \\
Jev & arm 2 & 30\% & 0.0757 (53) & 0.1786 (125) & 1 & 1 \\
Haiku & arm 1 & 0\% & 0.1443 (101) & 0.1443 (101) & 0 & 0 \\
Haiku & arm 1 & 10\% & 0.1586 (111) & 0.1714 (120) & 0 & 0 \\
Haiku & arm 1 & 30\% & 0.1714 (120) & 0.2371 (166) & 1 & 1 \\
Haiku & arm 2 & 0\% & 0.1457 (102) & 0.1657 (116) & 0 & 0 \\
Haiku & arm 2 & 10\% & 0.1443 (101) & 0.1829 (128) & 1 & 1 \\
Haiku & arm 2 & 30\% & 0.1471 (103) & 0.2214 (155) & 1 & 1 \\
\end{longtable}

Counts are out of 700 test items per cell. At 30\% noise the picked reduct changed fewer verdicts than the other in all four registered comparisons (H2, Section 7.8). The unsafe columns are equal in every row, because every flip came from the shared approval field. E2 is a substitution test between two reducts with nested sensed-field sets. It is not a general selection experiment.

Only the 30\% bounds were registered. The table below gives post hoc split-B bounds \(B^{+}\) at every noise level, computed the same way but not registered, beside the flips per reduct. They are descriptive.

\begin{longtable}[]{@{}
  >{\raggedright\arraybackslash}p{(\columnwidth - 12\tabcolsep) * \real{0.0920}}
  >{\raggedright\arraybackslash}p{(\columnwidth - 12\tabcolsep) * \real{0.1034}}
  >{\raggedright\arraybackslash}p{(\columnwidth - 12\tabcolsep) * \real{0.0920}}
  >{\raggedright\arraybackslash}p{(\columnwidth - 12\tabcolsep) * \real{0.2414}}
  >{\raggedright\arraybackslash}p{(\columnwidth - 12\tabcolsep) * \real{0.0920}}
  >{\raggedright\arraybackslash}p{(\columnwidth - 12\tabcolsep) * \real{0.0920}}
  >{\raggedright\arraybackslash}p{(\columnwidth - 12\tabcolsep) * \real{0.2874}}@{}}
\toprule\noalign{}
\begin{minipage}[b]{\linewidth}\raggedright
sensor
\end{minipage} & \begin{minipage}[b]{\linewidth}\raggedright
arm
\end{minipage} & \begin{minipage}[b]{\linewidth}\raggedright
reduct
\end{minipage} & \begin{minipage}[b]{\linewidth}\raggedright
post hoc \(B^{+}\), 0\%
\end{minipage} & \begin{minipage}[b]{\linewidth}\raggedright
10\%
\end{minipage} & \begin{minipage}[b]{\linewidth}\raggedright
30\%
\end{minipage} & \begin{minipage}[b]{\linewidth}\raggedright
flips at 0\% / 10\% / 30\%
\end{minipage} \\
\midrule\noalign{}
\endhead
\bottomrule\noalign{}
\endlastfoot
Jev & arm 1 & \(R_{\mathrm{branch}}\) & 0.0395 & 0.0395 & 0.0726 & 0 / 0 / 1 \\
Jev & arm 1 & \(R_{\mathrm{env}}\) & 0.0198 & 0.0198 & 0.0414 & 0 / 0 / 1 \\
Jev & arm 2 & \(R_{\mathrm{branch}}\) & 0.0198 & 0.0198 & 0.0198 & 0 / 1 / 1 \\
Jev & arm 2 & \(R_{\mathrm{env}}\) & 0.0395 & 0.0395 & 0.0510 & 0 / 1 / 1 \\
Haiku & arm 1 & \(R_{\mathrm{branch}}\) & 0.0395 & 0.0611 & 0.1623 & 0 / 0 / 1 \\
Haiku & arm 1 & \(R_{\mathrm{env}}\) & 0.0198 & 0.0198 & 0.1024 & 0 / 0 / 1 \\
Haiku & arm 2 & \(R_{\mathrm{branch}}\) & 0.0198 & 0.0509 & 0.0774 & 0 / 1 / 1 \\
Haiku & arm 2 & \(R_{\mathrm{env}}\) & 0.0395 & 0.1018 & 0.1716 & 0 / 1 / 1 \\
\end{longtable}

\hypertarget{e3-admission-ablation}{%
\subsection{E3: admission ablation}\label{e3-admission-ablation}}

E3 re-reads the E1 caches; no new call was made. Coverage is the fraction of test items decided. The unsafe rate is over all test items. Exposure is the verdict-change rate among decided items.

\begin{longtable}[]{@{}
  >{\raggedright\arraybackslash}p{(\columnwidth - 12\tabcolsep) * \real{0.1216}}
  >{\raggedright\arraybackslash}p{(\columnwidth - 12\tabcolsep) * \real{0.1351}}
  >{\raggedright\arraybackslash}p{(\columnwidth - 12\tabcolsep) * \real{0.1081}}
  >{\raggedright\arraybackslash}p{(\columnwidth - 12\tabcolsep) * \real{0.0946}}
  >{\raggedright\arraybackslash}p{(\columnwidth - 12\tabcolsep) * \real{0.1622}}
  >{\raggedright\arraybackslash}p{(\columnwidth - 12\tabcolsep) * \real{0.1622}}
  >{\raggedright\arraybackslash}p{(\columnwidth - 12\tabcolsep) * \real{0.2162}}@{}}
\toprule\noalign{}
\begin{minipage}[b]{\linewidth}\raggedright
ceiling
\end{minipage} & \begin{minipage}[b]{\linewidth}\raggedright
unknown
\end{minipage} & \begin{minipage}[b]{\linewidth}\raggedright
sensor
\end{minipage} & \begin{minipage}[b]{\linewidth}\raggedright
noise
\end{minipage} & \begin{minipage}[b]{\linewidth}\raggedright
coverage
\end{minipage} & \begin{minipage}[b]{\linewidth}\raggedright
unsafe rate
\end{minipage} & \begin{minipage}[b]{\linewidth}\raggedright
exposure among decided
\end{minipage} \\
\midrule\noalign{}
\endhead
\bottomrule\noalign{}
\endlastfoot
1\% & deny & Jev & 0\% & 1.0000 & 0.0000 & 0.0000 \\
1\% & deny & Jev & 10\% & 1.0000 & 0.0000 & 0.0429 \\
1\% & deny & Jev & 30\% & 1.0000 & 0.0014 & 0.1071 \\
1\% & deny & Haiku & 0\% & 1.0000 & 0.0000 & 0.0643 \\
1\% & deny & Haiku & 10\% & 1.0000 & 0.0000 & 0.0943 \\
1\% & deny & Haiku & 30\% & 1.0000 & 0.0000 & 0.1471 \\
1\% & escalate & Jev & 0\% & 1.0000 & 0.0000 & 0.0000 \\
1\% & escalate & Jev & 10\% & 0.8457 & 0.0000 & 0.0034 \\
1\% & escalate & Jev & 30\% & 0.6000 & 0.0014 & 0.0167 \\
1\% & escalate & Haiku & 0\% & 0.7557 & 0.0000 & 0.0000 \\
1\% & escalate & Haiku & 10\% & 0.6243 & 0.0000 & 0.0000 \\
1\% & escalate & Haiku & 30\% & 0.4086 & 0.0000 & 0.0175 \\
5\% & deny & Jev & 0\% & 1.0000 & 0.0000 & 0.0000 \\
5\% & deny & Jev & 10\% & 1.0000 & 0.0000 & 0.0286 \\
5\% & deny & Jev & 30\% & 1.0000 & 0.0014 & 0.0914 \\
5\% & deny & Haiku & 0\% & 1.0000 & 0.0000 & 0.0429 \\
5\% & deny & Haiku & 10\% & 1.0000 & 0.0000 & 0.0643 \\
5\% & deny & Haiku & 30\% & 1.0000 & 0.0000 & 0.1271 \\
5\% & escalate & Jev & 0\% & 1.0000 & 0.0000 & 0.0000 \\
5\% & escalate & Jev & 10\% & 0.8843 & 0.0000 & 0.0048 \\
5\% & escalate & Jev & 30\% & 0.6700 & 0.0014 & 0.0235 \\
5\% & escalate & Haiku & 0\% & 0.8457 & 0.0000 & 0.0000 \\
5\% & escalate & Haiku & 10\% & 0.7200 & 0.0000 & 0.0020 \\
5\% & escalate & Haiku & 30\% & 0.5029 & 0.0000 & 0.0256 \\
\end{longtable}

Routing unknown to escalation instead of deny trades coverage for exposure among decided items: fewer items are decided automatically, and in every cell where any verdict changed, the decided items changed verdict less often. Raising the ceiling from 1\% to 5\% raised coverage under escalation in every cell where it was below one. The unsafe rate did not move between the two ceilings. These readings are descriptive; E3 registered no hypothesis.

\hypertarget{e4-witness-and-implementation-check}{%
\subsection{E4: witness and implementation check}\label{e4-witness-and-implementation-check}}

E4 runs the two N6 witnesses through the experiment pipeline, with no model.

\begin{longtable}[]{@{}
  >{\raggedright\arraybackslash}p{(\columnwidth - 8\tabcolsep) * \real{0.2326}}
  >{\raggedright\arraybackslash}p{(\columnwidth - 8\tabcolsep) * \real{0.3488}}
  >{\raggedright\arraybackslash}p{(\columnwidth - 8\tabcolsep) * \real{0.1395}}
  >{\raggedright\arraybackslash}p{(\columnwidth - 8\tabcolsep) * \real{0.1163}}
  >{\raggedright\arraybackslash}p{(\columnwidth - 8\tabcolsep) * \real{0.1628}}@{}}
\toprule\noalign{}
\begin{minipage}[b]{\linewidth}\raggedright
witness
\end{minipage} & \begin{minipage}[b]{\linewidth}\raggedright
measured deny-to-allow (95\% CI)
\end{minipage} & \begin{minipage}[b]{\linewidth}\raggedright
checker's exact rate
\end{minipage} & \begin{minipage}[b]{\linewidth}\raggedright
S1 bound
\end{minipage} & \begin{minipage}[b]{\linewidth}\raggedright
pipeline matches checker
\end{minipage} \\
\midrule\noalign{}
\endhead
\bottomrule\noalign{}
\endlastfoot
(a) disjoint errors & 0.0452 {[}0.0412, 0.0495{]} (452/10000) & 1/20 & 1/20 & yes \\
(b) joint errors, AND contract & 0.0503 {[}0.0461, 0.0548{]} (503/10000) & 1/20 & 1/10 & yes \\
\end{longtable}

The registered check compares the pipeline's totals with the checker's own simulation, and it passed for both witnesses. A stricter comparison of the per-item outcome vectors, item by item over the 10,000 draws, found them equal to the checker's. Both paths evaluate the contract with the same evaluator, \texttt{sensed\_authority.bound.ContractModel}, so the check shows that the pipeline's sampling and bookkeeping agree with the checker's; it is not an independent test of contract evaluation. In witness (b) the S1 bound is twice the exact rate and holds; the v1 directional bound, 0, lies below the exact rate and fails. One reading needs care. In witness (a) the exact rate lies just above the simulation's 95\% interval. That is a property of one seeded draw of 10,000 items, not of the pipeline: the checker's own simulation, with the same seed, gives the same 452 flips.

\hypertarget{e5-calibration}{%
\subsection{E5: calibration}\label{e5-calibration}}

Calibration is measured on the test-split scores of every field and value question, against the assertion label.

\begin{longtable}[]{@{}
  >{\raggedright\arraybackslash}p{(\columnwidth - 12\tabcolsep) * \real{0.1233}}
  >{\raggedright\arraybackslash}p{(\columnwidth - 12\tabcolsep) * \real{0.1233}}
  >{\raggedright\arraybackslash}p{(\columnwidth - 12\tabcolsep) * \real{0.1370}}
  >{\raggedright\arraybackslash}p{(\columnwidth - 12\tabcolsep) * \real{0.1918}}
  >{\raggedright\arraybackslash}p{(\columnwidth - 12\tabcolsep) * \real{0.1507}}
  >{\raggedright\arraybackslash}p{(\columnwidth - 12\tabcolsep) * \real{0.1370}}
  >{\raggedright\arraybackslash}p{(\columnwidth - 12\tabcolsep) * \real{0.1370}}@{}}
\toprule\noalign{}
\begin{minipage}[b]{\linewidth}\raggedright
domain
\end{minipage} & \begin{minipage}[b]{\linewidth}\raggedright
sensor
\end{minipage} & \begin{minipage}[b]{\linewidth}\raggedright
answers
\end{minipage} & \begin{minipage}[b]{\linewidth}\raggedright
Spiegelhalter \(Z\)
\end{minipage} & \begin{minipage}[b]{\linewidth}\raggedright
\(p\)
\end{minipage} & \begin{minipage}[b]{\linewidth}\raggedright
Brier
\end{minipage} & \begin{minipage}[b]{\linewidth}\raggedright
BSS
\end{minipage} \\
\midrule\noalign{}
\endhead
\bottomrule\noalign{}
\endlastfoot
CH-C1 & Jev & 21000 & -40.5181 & 0.00e+00 & 0.0065 & 0.9566 \\
CH-C1 & Haiku & 21000 & 7.1036 & 1.22e-12 & 0.0142 & 0.9058 \\
CH-B1 & Jev & 21000 & -48.8080 & 0.00e+00 & 0.0113 & 0.9516 \\
CH-B1 & Haiku & 21000 & -3.6214 & 0.0003 & 0.0315 & 0.8655 \\
\end{longtable}

Both sensor families were informative, not calibrated. The Brier skill score \citep{brier1950verification} ranged from 0.8655 to 0.9566, far above zero, the score of a forecast that always gives the base rate. The Spiegelhalter test \citep{spiegelhalter1986probabilistic} rejected calibration in all four cases (largest \(p\) = 0.0003). A high score therefore does not mean an equally high chance that the document says so \citep{guo2017calibration}. This is why the thresholds are fitted on held-out data and not read off the score.

\hypertarget{hypotheses}{%
\subsection{Hypotheses}\label{hypotheses}}

The registered family is H1 and the four H2 tests, with Holm's correction \citep{holm1979simple} at \(\alpha\) = 0.05. H1 combines the 6 E1 cells' exact one-sided binomial tests with a Bonferroni factor. H2 uses an exact McNemar test \citep{mcnemar1947note} on the paired outcomes at 30\% noise. Intervals for H2 come from an item-clustered bootstrap with 10000 resamples (seed 20261104). The table is as in \texttt{results/p6.md}.

\begin{longtable}[]{@{}
  >{\raggedright\arraybackslash}p{(\columnwidth - 10\tabcolsep) * \real{0.2456}}
  >{\raggedright\arraybackslash}p{(\columnwidth - 10\tabcolsep) * \real{0.2105}}
  >{\raggedright\arraybackslash}p{(\columnwidth - 10\tabcolsep) * \real{0.1404}}
  >{\raggedright\arraybackslash}p{(\columnwidth - 10\tabcolsep) * \real{0.1140}}
  >{\raggedright\arraybackslash}p{(\columnwidth - 10\tabcolsep) * \real{0.1140}}
  >{\raggedright\arraybackslash}p{(\columnwidth - 10\tabcolsep) * \real{0.1754}}@{}}
\toprule\noalign{}
\begin{minipage}[b]{\linewidth}\raggedright
hypothesis
\end{minipage} & \begin{minipage}[b]{\linewidth}\raggedright
estimate
\end{minipage} & \begin{minipage}[b]{\linewidth}\raggedright
95\% CI
\end{minipage} & \begin{minipage}[b]{\linewidth}\raggedright
\(p\)
\end{minipage} & \begin{minipage}[b]{\linewidth}\raggedright
Holm-adjusted \(p\)
\end{minipage} & \begin{minipage}[b]{\linewidth}\raggedright
verdict
\end{minipage} \\
\midrule\noalign{}
\endhead
\bottomrule\noalign{}
\endlastfoot
H1: no E1 cell's unsafe rate exceeds its split-B bound (Bonferroni over 6 cells) & see E1 table & & 1.0000 & 1.0000 & not refuted: no violation detected \\
H2, Jev, arm 1: picked reduct has lower exposure at 30\% noise & -0.1000 (b = 0, c = 70, n = 700) & {[}-0.1229, -0.0786{]} & 1.69e-21 & 6.78e-21 & supported: picked reduct lower \\
H2, Jev, arm 2: picked reduct has lower exposure at 30\% noise & -0.1029 (b = 0, c = 72, n = 700) & {[}-0.1257, -0.0814{]} & 4.24e-22 & 2.12e-21 & supported: picked reduct lower \\
H2, Claude Haiku 4.5, arm 1: picked reduct has lower exposure at 30\% noise & -0.0657 (b = 0, c = 46, n = 700) & {[}-0.0843, -0.0486{]} & 2.84e-14 & 5.68e-14 & supported: picked reduct lower \\
H2, Claude Haiku 4.5, arm 2: picked reduct has lower exposure at 30\% noise & -0.0743 (b = 0, c = 52, n = 700) & {[}-0.0943, -0.0557{]} & 4.44e-16 & 1.33e-15 & supported: picked reduct lower \\
\end{longtable}

S2 orders the estimated bounds of nested contracts. It does not say that an item whose verdict changes under the smaller sensed set also changes under the larger one. A zero b in all four tests (b = 0, 0, 0 and 0) is therefore an empirical observation, not a consequence of S2: with reachable tuples on which z equals y and a verdict that allows if and only if x equals y, the contracts \{x, z\} and \{x, y\} are both sufficient, and at a true tuple with all three fields false, one joint misreading of x and y as true changes the verdict of the first, which senses only x, but not of the second.

``Not refuted'' for H1 means no violation was detected on 700 items per cell. It is never presented as evidence that the bound holds in general. Everything outside this table is descriptive.

\hypertarget{related-work-and-fence}{%
\section{Related work and fence}\label{related-work-and-fence}}

\textbf{Not claimed as new.} Thresholding a score with a reject option goes back to Chow \citep{chow1970optimum} and is studied for deep networks as selective classification \citep{geifman2017selective}. Conformal prediction and risk control give distribution-free coverage for set-valued predictions \citep{vovk2022algorithmic, angelopoulos2023gentle}. Learning to defer trains a model to hand cases to a human \citep{madras2018predict, mozannar2020consistent}. Calibration measurement is standard \citep{spiegelhalter1986probabilistic, brier1950verification, guo2017calibration}. Confidence-scored information extraction and human-in-the-loop escalation are common practice. Runtime enforcement and shields were fenced in paper 4, and reduct theory and cost-sensitive reduction \citep{pawlak1982rough} in paper 5. Closest to the selection component, Zhao, Min and Zhu choose test-cost-sensitive reducts for data with normally distributed measurement errors, under a constraint that preserves decision information \citep{zhao2013test}; there, measurement error enters reduct selection, but nothing abstains, no verdict-exposure bound is stated, and no sufficiency-checked authority contract decides.

\textbf{Prior compositions.} Composing an uncertain sensing component with a symbolic check that decides has been done before. Zhu and Zhang add probabilistic attributes, inferred by learned models, to access policies with automated reasoning support; their location example obtains an access-control attribute through probabilistic sensing \citep{zhu2024probabilistic}. Astorga et al.~synthesise perception contracts and check that they preserve a downstream invariant \citep{astorga2023perception}. Artikis et al.~threshold uncertain observations and process the admitted events deterministically \citep{artikis2012event}. The composition of a sensor with a deciding check is therefore not claimed as new.

\textbf{Union bounds.} Barthe et al.~give a program logic whose judgements carry a failure budget that adds under composition \citep{barthe2016program}. Assigning each sensed field the assertion ``correct and definite'' and using sufficiency to imply verdict agreement instantiates that reasoning; S1 restricts the sum to the contract's sensed fields. Waite et al.~combine perception-error union bounds with deterministic verification of autonomous systems \citep{waite2025state}. Neither states a bound indexed by an authority contract's sensed fields, but both occupy the broader statistical-to-symbolic composition.

\textbf{What is claimed} is the conjunction: an abstaining admission policy between a sensor and a sufficiency-checked authority contract; an exposure bound stated over that contract's sensed fields; and sensing-aware selection among sufficient contracts as a paper 5 cost model, for nested sensed-field sets. Neither the admission policy alone, nor the bound alone, nor the composition with a symbolic decision layer alone is claimed. This paper does not introduce abstention, calibration, conformal control or reducts. It shows how an abstaining sensor composes with a sufficiency-checked contract, bounds the resulting exposure by the contract's sensed fields, and selects contracts to minimise its estimated bound.

\hypertarget{claims-versus-non-claims}{%
\section{Claims versus non-claims}\label{claims-versus-non-claims}}

Each claim's status is taken from \texttt{CLAIMS.md}, where the evidence column names the slot or checker output behind it.

\begin{longtable}[]{@{}
  >{\raggedright\arraybackslash}p{(\columnwidth - 6\tabcolsep) * \real{0.0625}}
  >{\raggedright\arraybackslash}p{(\columnwidth - 6\tabcolsep) * \real{0.5833}}
  >{\raggedright\arraybackslash}p{(\columnwidth - 6\tabcolsep) * \real{0.1667}}
  >{\raggedright\arraybackslash}p{(\columnwidth - 6\tabcolsep) * \real{0.1875}}@{}}
\toprule\noalign{}
\begin{minipage}[b]{\linewidth}\raggedright
claim
\end{minipage} & \begin{minipage}[b]{\linewidth}\raggedright
statement
\end{minipage} & \begin{minipage}[b]{\linewidth}\raggedright
type
\end{minipage} & \begin{minipage}[b]{\linewidth}\raggedright
status
\end{minipage} \\
\midrule\noalign{}
\endhead
\bottomrule\noalign{}
\endlastfoot
C1 & A sensed field enters the gate only as an observation record with provenance, score and admission stamp; the model never emits a verdict. & architecture & supported \\
C2 & Proposition S1 bounds the probability of a verdict change by the sum of the sensed fields' wrong and unknown rates, and of a deny-to-allow change by the sum of their wrong rates. & instantiation & instantiation \\
C3 & For nested sensed-field sets, the bound does not increase as sensed fields are removed; minimising the estimated bound is a paper 5 cost model. & formal & supported (finite models, nested sets) \\
C4 & In every cell with a registered bound (all of E1; E2 at 30\% noise), with two sensor families, the observed unsafe rate does not exceed it. & empirical, registered & not refuted \\
C5 & Deny-to-allow flips were rare (13 of 21,000 test verdicts, each 1 in 700) and each one in a cell with a registered bound lay under it. & empirical, registered & supported \\
C6 & Scores from both sensor families are not calibrated, so admission thresholds must be set empirically. & empirical & supported \\
C7 & Sensing-aware reduct selection changes the choice in arm 1 and coincides in arm 2, relative to paper 5's choice on CH-B1, and the picked reduct had lower measured exposure in all four tests. & empirical, registered & supported \\
C8 & Everything reproduces from committed caches; no number is hand-entered. & reproducibility & supported \\
\end{longtable}

The non-claims, verbatim from \texttt{CLAIMS.md}: ``no prevalence in live systems; no claim that sensing is safe in general; no claim about Jev or Haiku beyond the tested configurations; no claim that the bound is tight.''

Correctness is relative to the declared loss model, candidate representation and reachable states; all domains are constructed.

\hypertarget{limitations}{%
\section{Limitations}\label{limitations}}

\textbf{Constructed corpora.} All documents are synthetic, in templated English, one statement per sensed field. Validation and test phrase banks are disjoint, but the rest of the document is shared across splits: the header lines, the filler bank, the noise templates, the date set, the names and the layout. A sensor can benefit from surface regularities that carry over from validation to test. No prevalence claim is made.

\textbf{Loose bounds.} S1 is valid, not tight, and the estimated bounds inherit its looseness. Every \(B^{+}\) in E1 is at least 0.0395, while the largest observed unsafe rate is 0.0014. Where verdicts changed, \(B\) was 6.6 to 7.3 times the observed change rate. Three causes add up. The Clopper-Pearson floor charges every sensed field a cost even with no observed error. The union bound adds per-field terms that overlap in practice. And the truth label counts every faithful reading of a contradictory record as a wrong admission, although most such readings do not change the verdict.

\textbf{Power of H1.} H1 is a one-sided test against a loose bound, so it can detect only large violations. The table gives, per E1 cell, the fewest flips among the 700 test items whose rate would exceed the cell's \(B^{+}\). H1 can fail only under a violation of that size.

\begin{longtable}[]{@{}
  >{\raggedright\arraybackslash}p{(\columnwidth - 8\tabcolsep) * \real{0.1404}}
  >{\raggedright\arraybackslash}p{(\columnwidth - 8\tabcolsep) * \real{0.1228}}
  >{\raggedright\arraybackslash}p{(\columnwidth - 8\tabcolsep) * \real{0.1579}}
  >{\raggedright\arraybackslash}p{(\columnwidth - 8\tabcolsep) * \real{0.3860}}
  >{\raggedright\arraybackslash}p{(\columnwidth - 8\tabcolsep) * \real{0.1930}}@{}}
\toprule\noalign{}
\begin{minipage}[b]{\linewidth}\raggedright
sensor
\end{minipage} & \begin{minipage}[b]{\linewidth}\raggedright
noise
\end{minipage} & \begin{minipage}[b]{\linewidth}\raggedright
\(B^{+}\)
\end{minipage} & \begin{minipage}[b]{\linewidth}\raggedright
fewest flips in 700 above \(B^{+}\)
\end{minipage} & \begin{minipage}[b]{\linewidth}\raggedright
flips observed
\end{minipage} \\
\midrule\noalign{}
\endhead
\bottomrule\noalign{}
\endlastfoot
Jev & 0\% & 0.0395 & 28 & 0 \\
Jev & 10\% & 0.0611 & 43 & 0 \\
Jev & 30\% & 0.1056 & 74 & 1 \\
Haiku & 0\% & 0.0395 & 28 & 0 \\
Haiku & 10\% & 0.0798 & 56 & 0 \\
Haiku & 30\% & 0.0707 & 50 & 0 \\
\end{longtable}

\textbf{Transfer and simultaneity.} The estimated bounds hold for the test split only if the sensors' error rates there are no higher than on split B. Split B used the validation phrase bank and the test split a disjoint one, and nothing in the design ensures that transfer; the observed rates are consistent with it but do not establish it. The 95\% limits are marginal: the sums \(B^{+}\) and \(B\), and the family of cells, carry no joint confidence level. The Bonferroni-adjusted versions in Section 7.2 give a simultaneous level within each cell; no adjustment across cells is made. Bounds for E2 below 30\% noise are post hoc.

\textbf{Two sensor configurations.} One configuration of each sensor family was tested, through the vendors' interfaces as pinned, on the run dates recorded in the manifests. Nothing is claimed about other versions, prompts or models.

\textbf{Shared-field dependence in E2.} Both reducts of an arm read the same approval field. Their outcomes are paired and their flips coincide (Section 7.3). The paired test accounts for this; the count of flips does not measure independent events.

\textbf{Stamp enforcement.} Record binding and stamp enforcement (a record whose request, resource, sensor version or policy version does not match is rejected) are covered by unit tests only. No experiment presented a mismatched record.

\textbf{Proof status.} The general proofs in Appendix A are tagged pending-human-review. The checkers cover finite model families and the witnesses.

\hypertarget{reproducibility}{%
\section{Reproducibility}\label{reproducibility}}

\textbf{Siblings and pins.} \texttt{engines.lock} pins \texttt{sarc-authority-derivation} at \texttt{cfb321e} (CH-B1 and CH-C1), \texttt{dqSarc} at \texttt{db6c396}, and this repository's own Jev probe record at \texttt{e1fc75b} (tag \texttt{jev-probes-final}). A run aborts before any model call if a pin differs.

\textbf{Caches and slots.} The raw responses are committed append-only in \texttt{responses/p6-E1.jsonl} and \texttt{responses/p6-E2.jsonl}, with run manifests. \texttt{python\ -m\ experiments.analysis\_p6} reads only the caches, the frozen thresholds and the pinned siblings, and writes \texttt{results/p6.md} and \texttt{results/p6.slots.json}; two runs give identical bytes. This manuscript is filled from those slots, the checker outputs and the E0 slots at the pinned tag by \texttt{python\ paper/populate.py}; a lint rejects any typed numeral that is not a registered design parameter. A reproduction from the caches calls no model. An independent researcher reproduced it from a fresh clone of the \texttt{v1.0-rc2} release on macOS under Python \texttt{3.13}, outside the Python \texttt{3.11} environment in which the repository was developed and checked (repository issue \#2): both runs of the analysis matched the committed results byte for byte, and the PDF was not rebuilt.

\textbf{Checkers and mutation testing.} \texttt{make\ formal} runs the S1, S2, S3 and N6 checkers twice and requires identical output. The S1 checker evaluated 4472, 131072 and 115200 pointwise cases on the three finite model families. Mutation testing of \texttt{src/sensed\_authority} with mutmut 3.7.0 killed 544 of 575 mutants with 0 untested, a kill score of 0.946 against a threshold of 0.85.

\textbf{Review and re-registration.} The external fence review, its attestation, the superseded registration and the re-registration are all committed and tagged. Three review rounds then examined the manuscript: R1 by the author's adjudicating model session, R2 by an external model (the reviewer of the fence), and R3 a mechanical replication from a fresh clone, which recomputed every reported table independently of this repository's code. All three are committed unedited, with attestations, under \texttt{review-secondary/}. Departures are logged in \texttt{prereg/DEVIATIONS.md}; none was recorded for this paper.

\textbf{Standing policy.} The arm health rule, introduced after KI-1, is now standing policy for every registered model run in this programme: a smoke check on the first 20 validation records and a 5\% invalid ceiling on the validation split, both final.

\textbf{Build.} \texttt{make\ paper} regenerates the manuscript, the figure, the bibliography and the PDF. The canonical document toolchain is Pandoc \texttt{3.1.3} and Tectonic \texttt{0.17.0} with its \texttt{default\_bundle\_v33} bundle; under it the LaTeX source, the PDF and the arXiv archive regenerate byte for byte, and \texttt{make\ release-check} refuses another Pandoc minor version. \texttt{make\ release-check} runs the tests, the checkers, the analysis regeneration, the lints and the release gates.

\textbf{Archive.} The repository is archived at DOI 10.5281/zenodo.23224016; this version corresponds to tag v1.0.1.

\hypertarget{acknowledgements}{%
\section*{Acknowledgements}\label{acknowledgements}}
\addcontentsline{toc}{section}{Acknowledgements}

An external model review of the novelty fence was commissioned before registration, committed unedited and not applied; the author re-registered after reading it. Eduardo Arana, independent researcher, reproduced the cached analysis from a fresh environment (repository issue \#2) and had no role in developing the repository; named with permission. AI assistance was used for drafting and engineering, under the author's direction; the author is solely responsible for the content.

\appendix

\hypertarget{proofs-and-checker-provenance}{%
\section{Proofs and checker provenance}\label{proofs-and-checker-provenance}}

\textbf{Proof of S1.} Fix a request with true tuple \(t\) and let \(o\) be the observation vector: recorded fields at their recorded values, sensed fields at their admitted values or unknown, derived fields computed from these. Let \(W_i\) be the event that sensed field \(i\) is admitted with a wrong value and \(U_i\) the event that it is unknown. Suppose no \(W_i\) and no \(U_i\) occurs. Recorded fields are at their true values (Section 2), and each derivation is truth-preserving, so a correctly admitted assertion derives the true value of its property. Then \(o\) agrees with \(t\) on every field of \(C\). Since \(t \in R\) agrees with \(o\), the verdict from observations is \(g(t')\) for some \(t'\) that agrees with \(t\) on \(C\), and sufficiency gives \(g(t') = g(t)\). So the change event lies inside \(\bigcup_i (W_i \cup U_i)\). If the verdict from observations is allow, no field is unknown, because unknown denies. So the deny-to-allow event lies inside \(\bigcup_i W_i\). The union bound gives both inequalities. The inclusions hold at every pair \((t, o)\), and both sides are linear in the joint distribution of \((t, o)\), so the inequalities hold for every distribution, including any dependence between fields. \(\square\)

\textbf{Proof of S2.} Let \(F_s' \subseteq F_s\) be nested sensed-field sets read by the same sensors on the same documents, through the same admission policy and the same derivation. Under these premises each field's admitted value is the same under both contracts; with a different policy or derivation it need not be, so \(e_i\) and \(u_i\) are the same per field, and the readings of the smaller set are the restriction of the larger. The S1 bound is a sum of non-negative terms over the set, so removing fields cannot increase it. The estimated bound uses the same per-field split-B estimates, so it is monotone in the same way. For sets that are not nested the per-field terms differ, and Section 4 shows a counterexample to the fewest-fields rule. \(\square\)

\textbf{Proof of S3.} Suppose unknown routes to deny and the configuration is deny-ward. Take any true tuple \(t\) with \(g(t) = \mathit{deny}\) and any realised observation. If a sensed field is unknown, the verdict is deny. Otherwise every sensed field is either true or admitted with some wrong value, and the recorded fields agree with \(t\); by the deny-ward condition, applied jointly to all sensed fields, the verdict is deny. So no realisation turns deny into allow, and the probability is zero. \(\square\)

\textbf{N6 constructions.} Witness (a): fields \(f_1, f_2\), contract allow if and only if both are true; true values are (false, true) or (true, false) with equal probability; the sensor reads the item's false field as true with probability 0.05 and is otherwise correct, never unknown. Each field is wrong with probability 1/40, the S1 deny-to-allow bound is 1/20, and the exact rate is 1/20: the bound is attained (attained: yes). Witness (b): the same contract with true values (false, false); with probability \(q\) = 1/20 both fields are admitted as true, otherwise both are correct. Each field is wrong with probability 1/20; the S1 bound is 1/10; the exact rate is 1/20; the v1 directional bound is 0; the configuration is not deny-ward in the global sense (globally deny-ward: no).

\textbf{Checker provenance.} The checkers enumerate three finite model families, B12, B3 and T2. B12 includes every sufficient contract, the empty one among them, which is sufficient exactly when the verdict is constant on the reachable set. \texttt{s1\_check} verified the S1 inclusions at 4472, 131072 and 115200 pairs of tuple and observation (holds: yes). \texttt{s2\_check} verified S2 for nested sets at 12308, 702464 and 373248 cases (holds: yes) and recorded the counterexample of Section 4. \texttt{s3\_check} enumerated 564, 2048 and 2048 configurations, of which 412, 318 and 530 were deny-ward, and confirmed that deny-ward holds exactly when no pair turns deny into allow (holds: yes). On B12 it found 4 configurations that a field-by-field test would call deny-ward but that are not deny-ward globally. \texttt{n6\_witness} computes both witnesses exactly and by simulation (holds: yes). Each output records its statement, scope, inputs hash and provenance; the provenance helper is ported with attribution from \texttt{sarc-suite-one-pass} through \texttt{sarc-authority-derivation}.

\hypertarget{registered-materials}{%
\section{Registered materials}\label{registered-materials}}

Everything in this appendix is rendered from \texttt{experiments/constants\_p6.py}, which the test suite checks verbatim against \texttt{prereg/p6-v1.1.md}, and from the frozen threshold files.

\textbf{Jev questions.} One Noul question per candidate value; each asks what the record states.

\begin{longtable}[]{@{}
  >{\raggedright\arraybackslash}p{(\columnwidth - 6\tabcolsep) * \real{0.1800}}
  >{\raggedright\arraybackslash}p{(\columnwidth - 6\tabcolsep) * \real{0.2200}}
  >{\raggedright\arraybackslash}p{(\columnwidth - 6\tabcolsep) * \real{0.2200}}
  >{\raggedright\arraybackslash}p{(\columnwidth - 6\tabcolsep) * \real{0.3800}}@{}}
\toprule\noalign{}
\begin{minipage}[b]{\linewidth}\raggedright
field
\end{minipage} & \begin{minipage}[b]{\linewidth}\raggedright
value
\end{minipage} & \begin{minipage}[b]{\linewidth}\raggedright
question id
\end{minipage} & \begin{minipage}[b]{\linewidth}\raggedright
question (verbatim)
\end{minipage} \\
\midrule\noalign{}
\endhead
\bottomrule\noalign{}
\endlastfoot
\texttt{approval\_assertion} & \texttt{not\_recorded} & \texttt{Q\_appr\_none} & Does this record state that no approval has been recorded for this change? \\
\texttt{approval\_assertion} & \texttt{recorded} & \texttt{Q\_appr\_recorded} & Does this record state that an approval for this change has been recorded? \\
\texttt{data\_residency\_region} & \texttt{us} & \texttt{Q\_residency\_us} & Does this record state that the resource's data is stored in the United States? \\
\texttt{data\_residency\_region} & \texttt{eu} & \texttt{Q\_residency\_eu} & Does this record state that the resource's data is stored in the European Union? \\
\texttt{data\_residency\_region} & \texttt{apac} & \texttt{Q\_residency\_apac} & Does this record state that the resource's data is stored in the Asia-Pacific region? \\
\texttt{branch} & \texttt{main} & \texttt{Q\_branch\_main} & Does this record state that the change targets the main branch? \\
\texttt{branch} & \texttt{staging} & \texttt{Q\_branch\_staging} & Does this record state that the change targets the staging branch? \\
\texttt{branch} & \texttt{feature} & \texttt{Q\_branch\_feature} & Does this record state that the change targets a feature branch? \\
\texttt{environment} & \texttt{production} & \texttt{Q\_env\_production} & Does this record state that the change targets the production environment? \\
\texttt{environment} & \texttt{staging} & \texttt{Q\_env\_staging} & Does this record state that the change targets the staging environment? \\
\texttt{environment} & \texttt{development} & \texttt{Q\_env\_development} & Does this record state that the change targets the development environment? \\
\texttt{approval\_assertion} & \texttt{recorded:2026-06-01} & \texttt{Q\_appr\_20260601} & Does this record state that an approval for this change was recorded on 2026-06-01? \\
\texttt{approval\_assertion} & \texttt{recorded:2026-06-15} & \texttt{Q\_appr\_20260615} & Does this record state that an approval for this change was recorded on 2026-06-15? \\
\texttt{approval\_assertion} & \texttt{recorded:2026-07-01} & \texttt{Q\_appr\_20260701} & Does this record state that an approval for this change was recorded on 2026-07-01? \\
\texttt{approval\_assertion} & \texttt{recorded:2026-07-15} & \texttt{Q\_appr\_20260715} & Does this record state that an approval for this change was recorded on 2026-07-15? \\
\texttt{approval\_assertion} & \texttt{recorded:2026-08-01} & \texttt{Q\_appr\_20260801} & Does this record state that an approval for this change was recorded on 2026-08-01? \\
\texttt{approval\_assertion} & \texttt{recorded:2026-08-15} & \texttt{Q\_appr\_20260815} & Does this record state that an approval for this change was recorded on 2026-08-15? \\
\end{longtable}

\textbf{Claude Haiku 4.5 prompt.} Temperature zero, no system prompt, the assistant prefill \texttt{\{}, and \texttt{max\_tokens\ =\ 64}, or \texttt{max\_tokens\ =\ 128} for the CH-C1 approval field. The user message:

\begin{verbatim}
You are reading one change-management record. Report only what the record states: {task}. Reply with only a JSON object of the form {object}, where each number is your probability that the record states that value. No code fences. No text after the object.

Record:
<rendered record text>
\end{verbatim}

\begin{longtable}[]{@{}
  >{\raggedright\arraybackslash}p{(\columnwidth - 4\tabcolsep) * \real{0.2400}}
  >{\raggedright\arraybackslash}p{(\columnwidth - 4\tabcolsep) * \real{0.4200}}
  >{\raggedright\arraybackslash}p{(\columnwidth - 4\tabcolsep) * \real{0.3400}}@{}}
\toprule\noalign{}
\begin{minipage}[b]{\linewidth}\raggedright
field
\end{minipage} & \begin{minipage}[b]{\linewidth}\raggedright
\texttt{\{task\}} (verbatim)
\end{minipage} & \begin{minipage}[b]{\linewidth}\raggedright
\texttt{\{object\}} keys
\end{minipage} \\
\midrule\noalign{}
\endhead
\bottomrule\noalign{}
\endlastfoot
\texttt{approval\_assertion} (CH-C1) & whether it states that no approval has been recorded, or that an approval was recorded, and on which date & \texttt{p\_not\_recorded}, \texttt{p\_2026\_06\_01}, \texttt{p\_2026\_06\_15}, \texttt{p\_2026\_07\_01}, \texttt{p\_2026\_07\_15}, \texttt{p\_2026\_08\_01}, \texttt{p\_2026\_08\_15} \\
\texttt{data\_residency\_region} (CH-C1) & where it states the resource's data are stored: us (United States), eu (European Union) or apac (Asia-Pacific) & \texttt{p\_us}, \texttt{p\_eu}, \texttt{p\_apac} \\
\texttt{approval\_assertion} (CH-B1) & whether it states that an approval has been recorded or that no approval has been recorded & \texttt{p\_recorded}, \texttt{p\_not\_recorded} \\
\texttt{branch} (CH-B1) & which branch it states the change targets: main, staging or feature & \texttt{p\_main}, \texttt{p\_staging}, \texttt{p\_feature} \\
\texttt{environment} (CH-B1) & which environment it states the change targets: production, staging or development & \texttt{p\_production}, \texttt{p\_staging}, \texttt{p\_development} \\
\end{longtable}

\textbf{E1 statement banks (CH-C1).} \texttt{\{approver\}}, \texttt{\{doc\}} and \texttt{\{ref\}} are drawn per item; \texttt{\{d\}} is the item's approval date.

\begin{longtable}[]{@{}
  >{\raggedright\arraybackslash}p{(\columnwidth - 6\tabcolsep) * \real{0.1786}}
  >{\raggedright\arraybackslash}p{(\columnwidth - 6\tabcolsep) * \real{0.1429}}
  >{\raggedright\arraybackslash}p{(\columnwidth - 6\tabcolsep) * \real{0.3393}}
  >{\raggedright\arraybackslash}p{(\columnwidth - 6\tabcolsep) * \real{0.3393}}@{}}
\toprule\noalign{}
\begin{minipage}[b]{\linewidth}\raggedright
field
\end{minipage} & \begin{minipage}[b]{\linewidth}\raggedright
value
\end{minipage} & \begin{minipage}[b]{\linewidth}\raggedright
validation bank
\end{minipage} & \begin{minipage}[b]{\linewidth}\raggedright
test bank
\end{minipage} \\
\midrule\noalign{}
\endhead
\bottomrule\noalign{}
\endlastfoot
\texttt{approval\_assertion} & \texttt{recorded} & \texttt{Approval:\ granted\ by\ \{approver\}\ on\ \{d\}.} / \texttt{Change\ approved\ by\ \{approver\}\ on\ \{d\}.} / \texttt{CAB\ approval\ recorded\ for\ \{ref\}\ on\ \{d\}.} & \texttt{Signed\ off\ by\ \{approver\},\ dated\ \{d\}.} / \texttt{\{approver\}\ recorded\ an\ approval\ for\ this\ change\ on\ \{d\}.} / \texttt{Approval\ entry\ dated\ \{d\},\ approver\ \{approver\}.} \\
\texttt{approval\_assertion} & \texttt{not\_recorded} & \texttt{Approval:\ none\ recorded.} / \texttt{Awaiting\ approval;\ no\ approver\ has\ signed\ off.} / \texttt{No\ CAB\ approval\ is\ on\ record\ for\ this\ change.} & \texttt{No\ sign-off\ has\ been\ entered\ for\ this\ change.} / \texttt{The\ approval\ log\ for\ this\ change\ is\ empty.} / \texttt{Nobody\ has\ approved\ this\ change\ yet.} \\
\texttt{data\_residency\_region} & \texttt{us} & \texttt{Data\ residency:\ United\ States.} / \texttt{Customer\ data\ for\ this\ resource\ is\ stored\ in\ US\ data\ centres.} / \texttt{Residency\ agreement\ \{doc\}\ places\ the\ data\ in\ the\ United\ States.} & \texttt{Hosting\ region:\ US.} / \texttt{The\ data\ for\ this\ resource\ are\ kept\ in\ the\ United\ States.} / \texttt{Storage\ location\ per\ contract\ \{doc\}:\ United\ States.} \\
\texttt{data\_residency\_region} & \texttt{eu} & \texttt{Data\ residency:\ European\ Union.} / \texttt{Customer\ data\ for\ this\ resource\ is\ stored\ in\ EU\ data\ centres.} / \texttt{Residency\ agreement\ \{doc\}\ places\ the\ data\ in\ the\ European\ Union.} & \texttt{Hosting\ region:\ EU.} / \texttt{The\ data\ for\ this\ resource\ are\ kept\ in\ the\ European\ Union.} / \texttt{Storage\ location\ per\ contract\ \{doc\}:\ European\ Union.} \\
\texttt{data\_residency\_region} & \texttt{apac} & \texttt{Data\ residency:\ Asia-Pacific.} / \texttt{Customer\ data\ for\ this\ resource\ is\ stored\ in\ APAC\ data\ centres.} / \texttt{Residency\ agreement\ \{doc\}\ places\ the\ data\ in\ the\ Asia-Pacific\ region.} & \texttt{Hosting\ region:\ APAC.} / \texttt{The\ data\ for\ this\ resource\ are\ kept\ in\ the\ Asia-Pacific\ region.} / \texttt{Storage\ location\ per\ contract\ \{doc\}:\ Asia-Pacific.} \\
\end{longtable}

\textbf{E2 statement banks (CH-B1).}

\begin{longtable}[]{@{}
  >{\raggedright\arraybackslash}p{(\columnwidth - 6\tabcolsep) * \real{0.1786}}
  >{\raggedright\arraybackslash}p{(\columnwidth - 6\tabcolsep) * \real{0.1429}}
  >{\raggedright\arraybackslash}p{(\columnwidth - 6\tabcolsep) * \real{0.3393}}
  >{\raggedright\arraybackslash}p{(\columnwidth - 6\tabcolsep) * \real{0.3393}}@{}}
\toprule\noalign{}
\begin{minipage}[b]{\linewidth}\raggedright
field
\end{minipage} & \begin{minipage}[b]{\linewidth}\raggedright
value
\end{minipage} & \begin{minipage}[b]{\linewidth}\raggedright
validation bank
\end{minipage} & \begin{minipage}[b]{\linewidth}\raggedright
test bank
\end{minipage} \\
\midrule\noalign{}
\endhead
\bottomrule\noalign{}
\endlastfoot
\texttt{approval\_assertion} & \texttt{recorded} & \texttt{Approval:\ granted\ by\ \{approver\}\ on\ \{date\}.} / \texttt{Change\ approved\ by\ \{approver\}\ under\ \{ref\}.} / \texttt{CAB\ approval\ recorded\ for\ \{ref\}\ on\ \{date\}.} & \texttt{Signed\ off\ by\ \{approver\},\ dated\ \{date\}.} / \texttt{\{approver\}\ recorded\ an\ approval\ for\ this\ change.} / \texttt{Approval\ entry\ dated\ \{date\},\ approver\ \{approver\}.} \\
\texttt{approval\_assertion} & \texttt{not\_recorded} & \texttt{Approval:\ none\ recorded.} / \texttt{Awaiting\ approval;\ no\ approver\ has\ signed\ off.} / \texttt{No\ CAB\ approval\ is\ on\ record\ for\ this\ change.} & \texttt{No\ sign-off\ has\ been\ entered\ for\ this\ change.} / \texttt{The\ approval\ log\ for\ this\ change\ is\ empty.} / \texttt{Nobody\ has\ approved\ this\ change\ yet.} \\
\texttt{branch} & \texttt{main} & \texttt{Target\ branch:\ main.} / \texttt{The\ change\ merges\ into\ the\ main\ branch.} / \texttt{Branch\ under\ change:\ main\ (trunk).} & \texttt{Merge\ target:\ main.} / \texttt{This\ patch\ goes\ onto\ main.} / \texttt{Destination\ branch\ is\ main.} \\
\texttt{branch} & \texttt{staging} & \texttt{Target\ branch:\ staging.} / \texttt{The\ change\ merges\ into\ the\ staging\ branch.} / \texttt{Branch\ under\ change:\ staging.} & \texttt{Merge\ target:\ staging.} / \texttt{This\ patch\ goes\ onto\ the\ staging\ branch.} / \texttt{Destination\ branch\ is\ staging.} \\
\texttt{branch} & \texttt{feature} & \texttt{Target\ branch:\ a\ feature\ branch.} / \texttt{The\ change\ lands\ on\ a\ feature\ branch.} / \texttt{Branch\ under\ change:\ feature/\{slug\}.} & \texttt{Merge\ target:\ feature/\{slug\}.} / \texttt{This\ patch\ goes\ onto\ a\ feature\ branch.} / \texttt{Destination\ branch\ is\ a\ feature\ branch.} \\
\texttt{environment} & \texttt{production} & \texttt{Target\ environment:\ production.} / \texttt{This\ change\ deploys\ to\ the\ production\ environment.} / \texttt{Environment:\ prod\ (customer-facing).} & \texttt{Deploys\ to:\ production.} / \texttt{Rollout\ target\ is\ the\ production\ environment.} / \texttt{Runs\ against\ prod.} \\
\texttt{environment} & \texttt{staging} & \texttt{Target\ environment:\ staging.} / \texttt{This\ change\ deploys\ to\ the\ staging\ environment.} / \texttt{Environment:\ pre-production\ staging.} & \texttt{Deploys\ to:\ staging.} / \texttt{Rollout\ target\ is\ the\ staging\ environment.} / \texttt{Runs\ against\ staging.} \\
\texttt{environment} & \texttt{development} & \texttt{Target\ environment:\ development.} / \texttt{This\ change\ deploys\ to\ the\ development\ environment.} / \texttt{Environment:\ dev\ sandbox.} & \texttt{Deploys\ to:\ development.} / \texttt{Rollout\ target\ is\ the\ development\ environment.} / \texttt{Runs\ against\ dev.} \\
\end{longtable}

\textbf{Admission thresholds, E1.} Fitted on split A at the 1\% ceiling. ``Never true'' and ``never false'' mean that no candidate threshold met the ceiling, so the value is never admitted in that direction.

\begin{longtable}[]{@{}
  >{\raggedright\arraybackslash}p{(\columnwidth - 10\tabcolsep) * \real{0.2600}}
  >{\raggedright\arraybackslash}p{(\columnwidth - 10\tabcolsep) * \real{0.2600}}
  >{\raggedright\arraybackslash}p{(\columnwidth - 10\tabcolsep) * \real{0.1200}}
  >{\raggedright\arraybackslash}p{(\columnwidth - 10\tabcolsep) * \real{0.1200}}
  >{\raggedright\arraybackslash}p{(\columnwidth - 10\tabcolsep) * \real{0.1200}}
  >{\raggedright\arraybackslash}p{(\columnwidth - 10\tabcolsep) * \real{0.1200}}@{}}
\toprule\noalign{}
\begin{minipage}[b]{\linewidth}\raggedright
field
\end{minipage} & \begin{minipage}[b]{\linewidth}\raggedright
value
\end{minipage} & \begin{minipage}[b]{\linewidth}\raggedright
Jev \(\tau^{\mathrm{true}}\)
\end{minipage} & \begin{minipage}[b]{\linewidth}\raggedright
Jev \(\tau^{\mathrm{false}}\)
\end{minipage} & \begin{minipage}[b]{\linewidth}\raggedright
Haiku \(\tau^{\mathrm{true}}\)
\end{minipage} & \begin{minipage}[b]{\linewidth}\raggedright
Haiku \(\tau^{\mathrm{false}}\)
\end{minipage} \\
\midrule\noalign{}
\endhead
\bottomrule\noalign{}
\endlastfoot
\texttt{approval\_assertion} & \texttt{not\_recorded} & 0.28 & 0.29 & never true & never false \\
\texttt{approval\_assertion} & \texttt{recorded:2026-06-01} & 0.45 & 0.02 & 0.05 & 0.05 \\
\texttt{approval\_assertion} & \texttt{recorded:2026-06-15} & 0.95 & 0.02 & 0.02 & 0.05 \\
\texttt{approval\_assertion} & \texttt{recorded:2026-07-01} & 0.94 & 0.02 & 0.15 & 0.01 \\
\texttt{approval\_assertion} & \texttt{recorded:2026-07-15} & 0.64 & 0.02 & 0.50 & 0.01 \\
\texttt{approval\_assertion} & \texttt{recorded:2026-08-01} & 0.63 & 0.02 & 0.50 & 0.01 \\
\texttt{approval\_assertion} & \texttt{recorded:2026-08-15} & 0.95 & 0.02 & 0.50 & 0.01 \\
\texttt{data\_residency\_region} & \texttt{apac} & 0.12 & 0.02 & 0.05 & 0.05 \\
\texttt{data\_residency\_region} & \texttt{eu} & 0.52 & 0.52 & 0.30 & 0.30 \\
\texttt{data\_residency\_region} & \texttt{us} & 0.35 & 0.02 & 0.50 & 0.00 \\
\end{longtable}

\textbf{Admission thresholds, E2.}

\begin{longtable}[]{@{}
  >{\raggedright\arraybackslash}p{(\columnwidth - 10\tabcolsep) * \real{0.2600}}
  >{\raggedright\arraybackslash}p{(\columnwidth - 10\tabcolsep) * \real{0.2600}}
  >{\raggedright\arraybackslash}p{(\columnwidth - 10\tabcolsep) * \real{0.1200}}
  >{\raggedright\arraybackslash}p{(\columnwidth - 10\tabcolsep) * \real{0.1200}}
  >{\raggedright\arraybackslash}p{(\columnwidth - 10\tabcolsep) * \real{0.1200}}
  >{\raggedright\arraybackslash}p{(\columnwidth - 10\tabcolsep) * \real{0.1200}}@{}}
\toprule\noalign{}
\begin{minipage}[b]{\linewidth}\raggedright
field
\end{minipage} & \begin{minipage}[b]{\linewidth}\raggedright
value
\end{minipage} & \begin{minipage}[b]{\linewidth}\raggedright
Jev \(\tau^{\mathrm{true}}\)
\end{minipage} & \begin{minipage}[b]{\linewidth}\raggedright
Jev \(\tau^{\mathrm{false}}\)
\end{minipage} & \begin{minipage}[b]{\linewidth}\raggedright
Haiku \(\tau^{\mathrm{true}}\)
\end{minipage} & \begin{minipage}[b]{\linewidth}\raggedright
Haiku \(\tau^{\mathrm{false}}\)
\end{minipage} \\
\midrule\noalign{}
\endhead
\bottomrule\noalign{}
\endlastfoot
\texttt{approval\_assertion} & \texttt{not\_recorded} & 0.35 & 0.09 & 0.98 & 0.02 \\
\texttt{approval\_assertion} & \texttt{recorded} & 0.42 & 0.46 & 0.30 & 0.30 \\
\texttt{branch} & \texttt{feature} & 0.57 & 0.57 & 0.34 & 0.05 \\
\texttt{branch} & \texttt{main} & 0.21 & 0.21 & 0.50 & 0.33 \\
\texttt{branch} & \texttt{staging} & 0.20 & 0.20 & 0.35 & 0.35 \\
\texttt{environment} & \texttt{development} & 0.28 & 0.28 & 1.00 & 0.34 \\
\texttt{environment} & \texttt{production} & 0.07 & 0.40 & 0.95 & 0.33 \\
\texttt{environment} & \texttt{staging} & 0.39 & 0.39 & 0.40 & 0.25 \\
\end{longtable}

\bibliographystyle{plainnat}
\bibliography{refs}

\end{document}